\documentclass[fleqn,twoside]{article}
\usepackage[headings]{espcrc2}

\readRCS
$Id: espcrc2.tex,v 1.2 2004/02/24 11:22:11 spepping Exp $
\usepackage{graphicx}
  \graphicspath{{../figures/}}
\usepackage[figuresright]{rotating}

\title{Remarkable suppression of the $e^{+}e^{-}\rightarrow\pi^{+}\pi^{-}$
contribution error into muon $g-2$}

\author{E. Barto\v s\address[SAV]{Inst. Phys., Slovak Acad. Sci., D\'ubravsk\'a cesta 9, 845\,11 Bratislava, Slovak Republic},
        S. Dubni\v cka\addressmark[SAV], %
        A.-Z. Dubni\v ckov\'a\address{FMPI Comenius University, Mlynsk\'a dolina, 842\,48 Bratislava, Slovak Republic}
        and
        A. Liptaj\addressmark[SAV]}

\runtitle{Error reduction of the pion contribution into muon $g-2$}
\runauthor{E. Barto\v s et al.}

\begin{document}

\begin{abstract}
We use the unitary and analytic model of the pion electromagnetic
form factor in order to evaluate in the lowest order the $e^{+}e^{-}\rightarrow\pi^{+}\pi^{-}$
contribution into the muon magnetic anomaly. We demonstrate, that
this technique enables us to reduce the uncertainty of the theoretical
prediction importantly in comparison to usual approaches, where the
measured data are integrated directly.
\end{abstract}

\maketitle

\section{Introduction and Motivation}

The question of an experimental evidence of physics beyond the Standard
Model (SM) has been and still is an issue in particle physics. This
topic is especially popular nowadays, when new high-energy experiments
are being constructed and put in work. Regarding the SM as a low-energy
approximation of a more general theory, one can expect the new physics
to manifest itself in two ways: important effects in measurements
at high energies or small effects at relatively low energies. The
latter is the case of the muon magnetic anomaly, often referred to
as the muon $g-2$ problem. To answer the question, whether some new
physic plays a role in this case, one needs to achieve a high precision
in both, theoretical prediction and experimental measurement.

The muon gyromagnetic ratio relates the muon spin and magnetic momentum
$\overrightarrow{\mu}=g_{\mu}\left(e/2m_{\mu}\right)\overrightarrow{s}$
and the magnetic anomaly is defined by $a_{\mu}=\left(g_{\mu}-2\right)/2.$
The higher muon mass makes its magnetic momentum, in comparison to
the magnetic momentum of the electron, more sensitive to the non-perturbative
hadronic effects and thus can the $a_{\mu}$ determination serve as
a good test field of the theory of strong interactions. A high-precision
experimental result is available, the anomaly $a_{\mu}$ was measured
by the $g-2$ collaboration in the E821 experiment at BNL with unprecedent
precision of 0.7 ppm \cite{g-2}. A high-precision theoretical prediction
is available also, but only in the electromagnetic and electroweak
sector the accuracy is satisfactory. In case of the hadronic contribution
$a_{\mu}^{had}$ more precision is desirable, because this contribution
is the dominant source of the total uncertainty of the theoretical
prediction, which does not allow to conclude on possible physics beyond
the SM. Depending on the author and the analyzed data one can cite
different numbers. If one takes the values from the summary publication
\cite{konstanty} $a_{\mu}^{exp}=116\,592\,093\,(63)\times10^{-11}$
and $a_{\mu}^{th}=116\,591\,810\,(210)\times10^{-11}$ one observes
$\triangle a_{\mu}\approx1.3\sigma_{\triangle a_{\mu}}$, $\triangle a_{\mu}=\left|a_{\mu}^{th}-a_{\mu}^{exp}\right|$.

The hadronic part is dominated by the lowest order and this can be
calculated using the formula\begin{equation}
a_{\mu}^{had,LO}=\frac{1}{3}\left(\frac{\alpha}{\pi}\right)^{2}\int_{4m_{\pi}^{2}}^{\infty}\frac{ds}{s}K(s)R(s),\label{eq:a_mi_had_LO}\end{equation}
where $K(s)=\int_{0}^{1}dx[x^{2}(1-x)]/[x^{2}+(1-x)\frac{s}{m_{\mu}^{2}}]$
and $R(s)=\sigma_{LO}\left(e^{+}e^{-}\rightarrow had\right)/\frac{4\pi\alpha^{2}}{3s}$.
The most important contribution comes from the pion channel $a_{\mu}^{had,LO}(e^{+}e^{-}\rightarrow\pi^{+}\pi^{-})$,
which we study in this work. The data are used to evaluate the integral
up to few GeV, above a perturbative calculation is possible. In order
to compare to other authors we chose three different limits for the
upper integration limit ($3.24\:\mathrm{GeV^{2}}$, $2.0449\:\mathrm{GeV^{2}}$
and $0.8\:\mathrm{GeV^{2}}$). 

The expression (\ref{eq:a_mi_had_LO}) was up to now \cite{Davier,Hagiwara}
always evaluated only by the direct integration of the $\sigma_{LO}\left(\pi^{+}\pi^{-}\right)$
data (with small exceptions, see \cite{Yndurain}). Our motivation
and our aim is to demonstrate, that the use of an appropriate model
allows for a dramatic error reduction. The error reduction in this
domain is of a crucial importance, since there is an effort to decrease
even more the experimental error \cite{buduce_ami} and so, if one
wants to solve the $g-2$ puzzle, the theoretical precision will have
to follow.

\section{Unitary and Analytic Model of the Pion Electromagnetic Form Factor}

We base our evaluation of $a_{\mu}^{had,LO}\left(\pi^{+}\pi^{-}\right)$
on the unitary and analytic (U\&A) model of the pion electromagnetic
(EM) form factor and relate the form factor to the cross section $\sigma_{tot}(e^{+}e^{-}\rightarrow\pi^{+}\pi^{-})=\frac{\pi\alpha^{2}}{3t}\left(1-4m_{\pi}^{2}/t\right)^{\frac{3}{2}}\left|F_{\pi}(t)\right|^{2}$,
so as to obtain $R(s)$. For a reliable prediction one needs to construct
a model that incorporates all known properties of the pion EM form
factor. The form factor is charge-normalized $F_{\pi}(0)=1$ and pQCD
predicts its asymptotic behavior $F_{\pi}(t)_{\left|t\right|\rightarrow\infty}\sim t^{-1}$,
where $t=q^{2}=-Q^{2}$. It is known, that $F_{\pi}(t)$ is an analytic
function in the whole complex $t$ plane, besides the branch points
and the cut on the positive real axis, the cut going from the lowest
branch point $t_{0}=4m_{\pi}^{2}$ up to $+\infty$. The pion form
factor has also cut on the second, unphysical Riemann sheet for $-\infty<t<0$
and satisfies the unitarity condition $Im\left[F_{\pi}(t)\right]=\left[A_{1}^{1}(t)\right]^{*}F_{\pi}(t)+\sigma(t)$,
where $A_{1}^{1}(t)$ is the \emph{P}-wave isovector transition amplitude
for elastic $\pi^{+}\pi^{-}$ scattering and $\sigma(t)$ approximates
all higher contributions. For $4m_{\pi}^{2}<t<(m_{\pi}+m_{\omega})^{2}$
one has $\sigma(t)=0$ and arrives to the so-called elastic unitarity
condition.

The construction of the U\&A model starts from the Vector Meson Dominance
(VMD) picture in which\[
F_{\pi}(t)=\sum_{v=\rho,\rho',\rho''}\frac{m_{v}^{2}}{m_{v}^{2}-t}\left(\frac{f_{v\pi\pi}}{f_{v}}\right),\]
where $f_{v\pi\pi}$ is the vector meson-pion coupling constant, $f_{v}$
is the universal vector meson coupling constant and three resonances
are taken into account. We proceed to a non-linear transformation
\[
t=t_{0}-\frac{4(t_{in}-t_{0})}{[1/W-W]^{2}}\]
in order to build in the cut, the lowest branch point $t_{0}$ and
an effective branch point $t_{in}$, the latter meant to approximate
all higher branch points. The model then becomes\[
F_{\pi}(W)=\left(\frac{1-W^{2}}{1-W_{N}^{2}}\right)^{2}\]
\[
\times\sum_{v=\rho,\rho',\rho''}\left[\frac{\left(W_{N}-W_{v0}\right)\left(W_{N}+W_{v0}\right)}{\left(W-W_{v0}\right)\left(W+W_{v0}\right)}\right.\]
\[
\left.\times\frac{\left(W_{N}-1/W_{v0}\right)\left(W_{N}+1/W_{v0}\right)}{\left(W-1/W_{v0}\right)\left(W+1/W_{v0}\right)}\left(f_{v\pi\pi}/f_{v}\right)\right],\]
where $W_{N}=W(t)|_{t=0}$ and $W_{v0}=W(t)|_{t=m_{v}^{2}}$. The
expression can be modified further; it can be shown that depending
on the relative positions of thresholds and masses one has $t_{0}<m_{v}<t_{in}\Rightarrow W_{v}=-W_{v}^{*}$
and $t_{in}<m_{v}\Rightarrow W_{v}=1/W_{v}^{*}$. The masses and $t_{in}$
are considered as free parameters of the model and so their relative
positions are not fixed. However, in order to provide explicit formulas
in the following text, we will here suppose $t_{0}<m_{\rho}<t_{in}<m_{\rho'},m_{\rho''}$,
what is actually being confirmed by the result of the data fitting.

\begin{figure}
\begin{centering}
\includegraphics[width=0.9\linewidth]{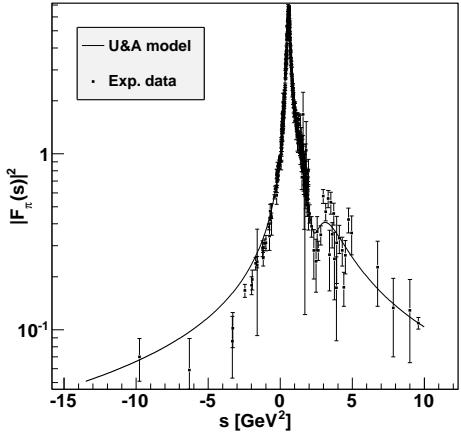}
\par\end{centering}

\caption{\label{fig:Fit}Fit of the data with the U\&A model.}

\end{figure}

The next step in the model construction is incorporation of the cut
on the second Riemann sheet. We use a succession of poles and zeros
generated by a rational function to approximate this cut (Pad\'{e}
approximation). We do it by adding to the model a multiplicative term
$\frac{\left(W-W_{Z}\right)\left(W_{N}-W_{P}\right)}{\left(W_{N}-W_{Z}\right)\left(W-W_{P}\right)}$,
where $W_{Z}$ and $W_{P}$ are free parameters from the interval
$0<W_{Z,P}<1$ (corresponds to $-\infty<t<0$). The construction of
the model is finally achieved by giving the vector mesons non-zero
decay widths $W_{v0}\rightarrow W_{v}=W(t)|_{t=\left(m_{v}-i\frac{\Gamma_{\nu}}{2}\right)^{2}}$.
The model gets form\[
F_{\pi}[W(t)]=\left(\frac{1-W^{2}}{1-W_{N}^{2}}\right)^{2}\frac{\left(W-W_{Z}\right)\left(W_{N}-W_{P}\right)}{\left(W_{N}-W_{Z}\right)\left(W-W_{P}\right)}\]

\[
\times\left[\frac{\left(W_{N}-W_{\rho}\right)\left(W_{N}-W_{\rho}^{*}\right)}{\left(W-W_{\rho}\right)\left(W-W_{\rho}^{*}\right)}\right.\]
\[
\times\frac{\left(W_{N}-1/W_{\rho}\right)\left(W_{N}-1/W_{\rho}^{*}\right)}{\left(W-1/W_{\rho}\right)\left(W-1/W_{\rho}^{*}\right)}\left(f_{\rho\pi\pi}/f_{\rho}\right)\]
\[
+\sum_{v=\rho',\rho''}\frac{\left(W_{N}-W_{v}\right)\left(W_{N}-W_{v}^{*}\right)}{\left(W-W_{v}\right)\left(W-W_{v}^{*}\right)}\]
\[
\left.\times\frac{\left(W_{N}+W_{v}\right)\left(W_{N}+W_{v}^{*}\right)}{\left(W+W_{v}\right)\left(W+W_{v}^{*}\right)}\left(f_{v\pi\pi}/f_{v}\right)\right].\]
The ratio of couplings $f_{\rho\pi\pi}/f_{\rho}$ is a free parameter
of the model, the two remaining ratios, $f_{\rho'\pi\pi}/f_{\rho}$
and $f_{\rho''\pi\pi}/f_{\rho''}$, can be related to the first one
by using the normalization condition $F_{\pi}(0)=1$ and by considering
the behavior of the imaginary part of the form factor at $q=0$. The
$\rho-\omega$ interference is taken into account when fitting the
experimental data by a Breit-Wigner term, the fitting function is\[
F_{\pi}\left[W(t)\right]+Re^{i\phi}\frac{m_{\omega}^{2}}{m_{\omega}^{2}-t-im_{\omega}\Gamma_{\omega}},\]
where $\phi=\arctan\frac{m_{\rho}\Gamma_{\rho}}{m_{\rho}^{2}-m_{\omega}^{2}}$
and the amplitude $R$ is considered as an additional free parameter.
The fitting function thus has 11 free parameters in total ($t_{in},$
three masses, three widths, ratio $f_{\rho\pi\pi}/f_{\rho}$, $W_{z}$,
$W_{P}$ and $R$) and is used to fit 523 experimental points measured
in experiments \emph{CLEO}, \emph{NA7}, \emph{OLYA}, \emph{CMD}, \emph{CMD-2},
\emph{SND}, \emph{KLOE}, in \emph{JINR} Dubna and in $F_{\pi}$ collaboration
at Jefferson Lab \cite{PiData_S01,PiData_S02,PiData_S03,PiData_S04,PiData_S05,PiData_S06,PiData_S07,PiData_S08,PiData_S09,PiData_S10,PiData_S11,PiData_S12,PiData_S13,PiData_S14,PiData_S15,PiData_S16,PiData_S17,PiData_S18,PiData_S19,PiData_S20,PiData_N1,PiData_N2,PiData_N3,PiData_N4,PiData_N5,PiData_N6,PiData_N7,PiData_N8}.
The best fit result is shown in Figure \ref{fig:Fit}.%

\section{Error Evaluation and Results}

Two approaches were used for the error evaluation. The first one was
a {}``Monte Carlo'' method based on a random number generator with
the assumption of the Gaussian distribution for the uncertainties
of the published data points. For each point a new one was randomly
generated using the Gaussian probability density function with the
mean identical to the original point and $\sigma$ equal to the published
error. Doing this for each point, new {}``random'' data set was
obtained. This data set was then fitted and the values of the parameters
of the model $p_{i}$ as well as the value of $a_{\mu}^{had,LO}\left(\pi^{+}\pi^{-}\right)$
were extracted. Repeating the whole procedure 4000 times, we reached
statistics high enough to allow us for a reliable error calculation.
The mean $\overline{a_{\mu}^{had,LO}}\left(\pi^{+}\pi^{-}\right)$
and the $\sigma$ were calculated from the 4000 values and since the
mean is not, in general, identical with the optimal-fit value $a_{\mu,OPT}^{had,LO}\left(\pi^{+}\pi^{-}\right)$
we present asymmetric uncertainties $a_{\mu}^{had,LO}\left(\pi^{+}\pi^{-}\right)=a_{\mu,OPT}^{had,LO}\left(\pi^{+}\pi^{-}\right)_{+B}^{-A}$,
where $A=\sigma+a_{\mu,OPT}^{had,LO}\left(\pi^{+}\pi^{-}\right)-\overline{a_{\mu}^{had,LO}}\left(\pi^{+}\pi^{-}\right)$
and $B=\sigma+\overline{a_{\mu}^{had,LO}}\left(\pi^{+}\pi^{-}\right)-a_{\mu,OPT}^{had,LO}\left(\pi^{+}\pi^{-}\right)$.

In the second approach the program MINUIT was used to establish the
uncertainties of the model parameters. Then, taking the numerical
derivatives for $\frac{\partial}{\partial p_{i}}a_{\mu}^{had,LO}\left(\pi^{+}\pi^{-}\right)$,
the uncertainty was propagated to $a_{\mu}^{had,LO}\left(\pi^{+}\pi^{-}\right)$
using the covariance matrix. In this method the errors are symmetric.

In addition to the integration of the model, we also performed a direct
integration of the data points based on the trapezoidal rule, so as
to cross-check our compatibility with other authors. Our results and
some results from other authors \cite{Davier,Hagiwara,Yndurain} are
summarized in Table \ref{tab:Results}. %
\begin{table*}
\begin{centering}
\begin{tabular}{llll}
\hline 
 & \multicolumn{3}{c}{$a_{\mu}^{had,LO}(e^{+}e^{-}\rightarrow\pi^{+}\pi^{-})\times10^{11}$}\tabularnewline
\cline{2-4} 
\noalign{\vskip\doublerulesep}
Interval $\left[GeV^{2}\right]$ & $4m_{\pi}^{2}<t<3.24$ & $4m_{\pi}^{2}<t<2.0449$ & $4m_{\pi}^{2}<t<0.8$\tabularnewline
\hline
\noalign{\vskip\doublerulesep}
Model, method 1 & $5132.36_{+0.83}^{-0.83}$ & $5128.22_{+0.73}^{-0.67}$ & $4870.24_{+0.20}^{-0.20}$\tabularnewline
\noalign{\vskip\doublerulesep}
Model, method 2 & $5132.37\pm3.00$ & $5128.25\pm2.86$ & $4870.44\pm2.64$\tabularnewline
\noalign{\vskip\doublerulesep}
Data & $5035.33_{+28.32}^{-17.22}$ & $5031.22_{+28.94}^{-16.43}$ & $4756.77_{+27.55}^{-18.14}$\tabularnewline
\noalign{\vskip\doublerulesep}
\noalign{\vskip\doublerulesep}
\emph{Davier} & $5040.00\pm31.05$ &  & \tabularnewline[\doublerulesep]
\noalign{\vskip\doublerulesep}
\emph{Hagiwara et al.} &  & $5008.2\pm28.70$ & \tabularnewline
\noalign{\vskip\doublerulesep}
\emph{Yndur\'{a}in et al.} &  &  & $4715\pm33.53$\tabularnewline
\hline
\noalign{\vskip\doublerulesep}
\end{tabular}
\par\end{centering}

\caption{\label{tab:Results}Our results and results of other authors \cite{Davier,Hagiwara,Yndurain}.}

\end{table*}

\section{Discussion, Summary and Outlook}

The use of the U\&A model dramatically reduces the error on $a_{\mu}^{had,LO}\left(\pi^{+}\pi^{-}\right)$.
This is not an arbitrary feature of the model but originates from
model-independent information which is additional to the data in the
integration region and which can be taken into account when the model
is used. The most important sources contributing to error reduction
are
\begin{itemize}
\item Expected smoothness of the $F_{\pi}(t)$ at small scale $\Delta t$:
The model provides a function behaving smoothly at small $\Delta t$.
\item Experimental data outside the integration region: The fit is done
not only to the data inside, but also to the data outside the integration
region.
\item Theoretical knowledge on $F_{\pi}(t)$: The model respects all known
properties of the pion EM form factor.
\end{itemize}
Especially the first point plays an important role. The new precise
data tend to lie above older, less precise data and, in some regions,
the vertical spread of the data is very important, at the limit of
inconsistency. If the calculation of the integral is based directly
on data, then less precise data shift the mean value of the integral
and enlarge the uncertainty. When the model is used, the predicted
behavior of $F_{\pi}(t)$ as given by the result of the fit is mostly
determined by precisely measured points and is only little influenced
by data with important uncertainties. This leads to more appropriate
mean value and smaller errors. In consequence, one arrives to the
mean value of $a_{\mu}^{had,LO}\left(\pi^{+}\pi^{-}\right)$ which
is higher then what is obtained by the direct data integration and
to much reduced uncertainty. The shift in the mean value goes in the
{}``right'' direction and brings the theoretical value closer to
the experimental one.

The error estimates from the two used methods are not fully compatible,
the first method gives smaller errors. This might be related to statistical
fluctuations (1st method) and to approximations - numerical derivatives
and linearization (2nd method).

In this article we presented the calculation of $a_{\mu}^{had,LO}\left(\pi^{+}\pi^{-}\right)$
based on the U\&A model. This approach allows for important error
reduction and we plan to use it also for evaluating the contributions
of other channels to $a_{\mu}^{had,LO}$.

\section*{Acknowledgments}

The work was partly supported by Slovak Grant Agency for Sciences
VEGA, grant No. 2/7116/29.

\end{document}